\documentclass[aps,preprint ,superscriptaddress,nofootinbib]{revtex4-2}
\usepackage{amsmath}
\allowdisplaybreaks[1]
\usepackage{mathtools}
\usepackage{physics}
\usepackage{graphicx}
\usepackage{amssymb}
\usepackage[colorlinks,citecolor=red]{hyperref}
\usepackage{slashed}
\usepackage{bm}
\usepackage{adjustbox}

\newcommand{\beq}{\begin{equation}}
	\newcommand{\eeq}{\end{equation}}

\newcommand\snowmass{\begin{center}\rule[-0.2in]{\hsize}{0.01in}\\\rule{\hsize}{0.01in}\\
		\vskip 0.1in Submitted to the  Proceedings of the US Community Study\\ 
		on the Future of Particle Physics (Snowmass 2021)\\ 
		\rule{\hsize}{0.01in}\\\rule[+0.1in]{\hsize}{0.01in} \end{center}}

\long\def\comment#1{}  

\usepackage{ulem}

\begin{document}
	
	\preprint{
		{\vbox {
				\hbox{\bf MSUHEP-22-015}
	}}}
	\vspace*{0.2cm}
	
\title{Impact of lattice $s(x) - \overline{s}(x)$ data in the CTEQ-TEA global analysis}

\author{Tie-Jiun Hou}
\affiliation{Center for Theory and Computation, National Tsing Hua University, Hsinchu 300, Taiwan}

\author{Huey-Wen Lin}
\affiliation{Department of Physics and Astronomy, Michigan State University, East Lansing, MI 48824}
\affiliation{Department of Computational Mathematics, Science \& Engineering, Michigan State University, East Lansing, MI 48824}

\author{Mengshi Yan}
\affiliation{Department of Physics and State Key Laboratory of Nuclear Physics and Technology, Peking University, Beijing 100871, China}

\author{C.-P. Yuan}
\affiliation{Department of Physics and Astronomy, Michigan State University, East Lansing, MI 48824}

\begin{abstract}
We study the impact of the lattice data on the determination of the strangeness asymmetry distribution $s_-(x) \equiv s(x) - {\bar s}(x)$ in the general CTEQ-TEA global analysis of parton distribution functions (PDFs) of the proton. Firstly, we find that allowing a non-vanishing $s_-(x)$, at the initial $Q_0=1.3$ GeV scale, in a global PDF analysis leads to the quality of the  CT18As fit similar to CT18A. Secondly, including the lattice data in the CT18As\_Lat fit greatly reduces the $s_-$-PDF error band size in the large $x$ region.
To further reduce its error would require more precise lattice data, extended to smaller $x$ values.
\end{abstract}
\date{\today}

\snowmass

\maketitle
\clearpage

\section{Introduction} \label{intro}

The Large Hadron Collider (LHC) has entered an era of precision physics. To match the experimental precision, it is necessary to have precise predictions in QCD theory, which require correspondingly precise parton distribution functions (PDFs), such as the recent CT18~\cite{Hou:2019efy}, MSHT20~\cite{Bailey:2020ooq} and NNPDF4.0~\cite{Ball:2021leu} PDFs obtained at the 
next-to-next-to-leading order (NNLO) accuracy in QCD.

In the CT18 analysis, the strange quark and anti-quark PDFs of the proton are assumed to be the same, {\it i.e.,} $\overline{s}(x)=s(x)$, at $Q_0=1.3$ GeV, where the nonperturbative PDFs are specified, and a non-vanishing $(s-\bar{s})(x)$ 
is generated at higher energy scales by DGLAP evolution~\cite{Moch:2004pa,Catani:2004nc}.
In~\cite{Hou:2019efy}, noticeable tensions between the original NuTev~\cite{Mason:2006qa} and CCFR~\cite{NuTeV:2001dfo} DIS dimuon data and the precision ATLAS $\sqrt{s} = 7$ TeV $W$, $Z$ data~\cite{ATLAS:2016nqi} were found. In MSHT20~\cite{Bailey:2020ooq}, it was concluded that allowing $s(x) \neq \bar{s}(x)$ at the $Q_0$ scale can release some of the above-mentioned tensions. 
In this work, we extend the CT18 analysis to allow a non-vanishing strangeness asymmetry, defined as 
$s_{-}(x) \equiv s(x) - \bar{s}(x)$, at the $Q_0$ scale, and the resulting PDF set is hereafter referred to as CT18As.

Besides the phenomenology approach of performing  global PDF analysis, 
a nonperturbative approach from first principles, such as lattice QCD (LQCD), provides hope to resolve many of the outstanding theoretical disagreements and provides information in regions that are unknown or difficult to observe in experiments.  
Recent breakthroughs, such as large-momentum effective theory (LaMET)~\cite{Ji:2013dva,Ji:2014gla,Ji:2020ect} (quasi-PDFs), has make it possible for lattice calculation to provide information on the $x$-dependent PDFs. There has been many pioneering works showing great promise in obtaining quantitative results for the unpolarized, helicity and transversity quark and antiquark distributions~\cite{Lin:2014gaa,Lin:2014yra,Lin:2014zya,Chen:2016utp,Alexandrou:2014pna,Alexandrou:2015rja}
using the quasi-PDFs approach~\cite{Ji:2013dva}. 
Increasingly many lattice works are being performed at physical pion mass since the first study in Ref.~\cite{Lin:2017ani}.
(A recent review of the theory and lattice calculations can be found in Refs.~\cite{Ji:2020ect,Lin:2017snn,Constantinou:2020hdm}.) 
The first  Bjorken-$x$--dependent of the strange PDF using lattice-QCD calculations was first reported in Ref.~\cite{Zhang:2020dkn}. 
The calculation is currently done using a single lattice spacing 0.12~fm with extrapolation to physical pion masses. 
In this work, we will use the extrapolated lattice matrix elements to calculate $s_{-}(x)$, which is then taken as an input {\it data} to further constrain $s(x)$ and $\bar{s}(x)$ at the $Q_0$ scale in the CT18-like global analysis, and the resulting PDF set is hereafter referred to as CT18As\_Lat. 
More details will be presented in the next section.

\section{Strangeness from Lattice and Global fitting} \label{fitting}

\subsection{Lattice data of strangeness asymmetry $s_{-}(x)$}
\label{fitting:svl-lat}

In this work, we took the LaMET coordinate-space matrix elements (extrapolated to physical pion mass) for strange PDF~\cite{Zhang:2020dkn} calculated on a 0.12~fm lattice spacing and $2+1+1$-flavor 310-MeV HISQ sea generated by MILC collaboration~\cite{MILC:2010pul,Bazavov:2012xda} at single lattice spacing. The matrix elements are renormalized 
nonperturbative renormalization (NPR) in RI/MOM scheme, the same strategy as in past works~\cite{Stewart:2017tvs,Chen:2017mzz}. 
Since $s_{-}(x)$ is flavor singlet, we can confidently calculate it using the LaMET matrix elements calculated in coordinate space on the lattice.  
We Fourier transform the renormalized matrix elements into quasi-PDFs by using the extrapolation formulation suggested by Ref.~\cite{Ji:2020brr} by fitting large-$|z|$ data using the formula $c_1(-izP_z)^{-d_1}+c_2 e^{izP_z}(izP_z)^{-d_2}$, inspired by the Regge behavior, to extrapolate the matrix elements into the region beyond the lattice calculation and suppress Fourier-transformation artifacts.  
The quasi-PDF can be related to the $P_z$-independent lightcone PDF through 
at scale $\mu$ in $\overline{\text{MS}}$ scheme through a factorization theorem~\cite{Ji:2014gla}
\begin{align}
	\label{eq:matching}
	\tilde{q}_\psi(x,&P_z,\mu^{\overline{\text{MS}}},\mu^\text{RI},p^\text{RI}_z)=\int_0^1 \frac{dy}{|y|} \times \nonumber\\ &C\left(\frac{x}{y},\left(\frac{\mu^\text{RI}}{p_z^\text{RI}}\right)^2,\frac{yP_z}{\mu^{\overline{\text{MS}}}},\frac{yP_z}{p_z^\text{RI}}\right)
	q_\psi(y,\mu^{\overline{\text{MS}}}) +...
\end{align}
where $p_z^\text{RI}$ and $\mu^\text{RI}$ are the momentum of the off-shell strange quark and the renormalization scale in the RI/MOM-scheme nonperturbative renormalization (NPR),
$C$ is a perturbative matching kernel used in our previous works~\cite{Chen:2018xof,Lin:2018qky,Chen:2018fwa,Chen:2019lcm}.
Note that the matching from quasi-PDF to PDF has residual  systematics at
$O\left(\frac{\Lambda_\text{QCD}^2}{(xP_z)^2}\right)$ and $O\left(\frac{\Lambda_\text{QCD}^2}{(1-x)^2P_z^2}\right)$ at very small $x$ and $x$ near 1. 
To take advantage of existing lattice data to reach a wider region of $x$, we choose to focus on $P_z \approx 1.7$~GeV.  
From the isovector nucleon PDF study, at this $P_z$ boost momentum, we can reasonably rely on lattice inputs for $x\in[0.3,0.8]$. 
Beyond this region, the lattice errors could increase significantly due to the systematics at finite momentum used here.


\subsection{Strangeness asymmetry $s_{-}(x)$ in CTEQ-TEA PDF analysis}
\label{fitting:svl-CT18}

In the nominal CTEQ-TEA PDF fitting \cite{Hou:2019efy, Dulat:2015mca, Lai:2010vv, Nadolsky:2008zw, Lai:2007dq}, the active parton flavours to be parametrised at $Q_0 = 1.3$ GeV are $u, \bar{u}, d, \bar{d}, s, \bar{s}, g$. In the parametrisation of sea quark distributions, $q_s(x, Q_0) = \bar{q}_s(x, Q_0)$ is imposed in the nominal CT PDFs. 
The CTEQ6.5S0 PDF~ \cite{Lai:2007dq} is dedicated as with focus on the strangeness sector, where the strangeness asymmetry $s_{-}(x, Q_0)$ is explicitly parametrised at $Q_0$.

In this work, we follow the strategy presented in \cite{Lai:2007dq}, but with updated experimental data and noperturbative parametrization forms of active partons at the $Q_0$ scale, together with NNLO theory predictions. More specifically, in the CT18As analysis, 
we start from the alternative PDF set, CT18A NNLO \cite{Hou:2019efy}, rather than the nominal CT18 NNLO fit. This is because the ATLAS $\sqrt{s} = 7$ TeV $W$, $Z$ combined cross section measurement \cite{ATLAS:2016nqi} (ID=248) data set is included in the CT18A fit, while it is absent the nominal CT18 fit.
In CT18 analysis, this data is found to prefer larger total strangeness, $s_{+} \equiv s(x) + {\bar s} (x)$, and to have tensions with other dimuon data \cite{Mason:2006qa, Mason:2006qa}, which is sensitive to strangeness distribution. 

CT18As fit adopts the same nonperturbative PDF forms as CT18A fit at the $Q_0$ scale, except strange quark and antiquark PDFs which are determined by $s_{+}(x)$ and $s_{-}(x)$. 
The parametrisation of the strange asymmetry distribution $s_{-}(x)$ should respect the number sum rule of the strangeness,
\begin{equation}
 \int_0^1 \ dx s_{-}(x) = 0.
 \label{eq:sv_nsr}
\end{equation}
In principle, a parametrisation with any number of crossing, with $s_{-}(x) = 0$, is possible, as long as Eq.~\eqref{eq:sv_nsr} is satisfied. Here, we focus on those parametrisation forms with only one crossing for $x$ between  $10^{-6}$ and 1.


To obtain CT18As\_Lat PDFs, we take 
the lattice data of strangeness asymmetry presented in Sec.~\ref{fitting:svl-lat} as a constraint to the global PDF fit, by using the Lagrange Multiplier method, as we have regarded the lattice results of $s_{-}(x)$ as an additional data on top of the data set of CT18A. Hence, CT18As\_Lat is the update of CT18As with the inclusion of the lattice data of $s_{-}(x)$ evaluated at the $Q_0$ scale.
\comment{
	The uncertainty of lattice calculation is treated as the uncorrelated error during the fitting. Therefore, the quality-of-fit receives the contribution of lattice calculation,
\begin{eqnarray}
 \chi^2 &=& \chi^2_{\text{Exp.}} + \chi^2_{\text{Lat.}} \nonumber \\
 &=& \chi^2_{\text{Exp.}} + \sum_i \Big{(} \frac{s_v^{\text{para.}}(x_i) - s_v^{\text{Lat.}}(x_i)}{ \Delta s_v^{\text{Lat.}}(x_i) } \Big{)}^2,
 \label{eq:chi2_LM}
\end{eqnarray}
where $\chi^2_{\text{Exp.}}$ is the total $\chi^2$ for fitting experimental data.
}


\section{Results} \label{results}

In this section, we discuss the quality of various fits and compare the resulting PDFs. 

The quality of the CT18A, CT18As, and CT18As\_Lat fits is compared in Table \ref{tab:quality}, which shows that they all have the same value of $\chi^2$ per number of data point, around 1.19. 
The difference of 10 units in  $\chi^2_\text{tot}$ is much smaller than the tolerance (with a difference of 100 units) used in the CT18 analysis to define the PDF uncertainty at the 90\% confidence level (CL).

\begin{table}
\begin{center}
\begin{tabular}{lccc}
PDF         & $s_{-}(x,Q_0)${  }  & Lat. data  & $\chi^2_\text{tot}$ \\
\hline
CT18A        & 0  & No & 4376 \\
CT18As       & $\neq 0$ & No  & 4344 \\
CT18As\_Lat  & $\neq 0$  &Yes  & 4363 \\
\hline
\end{tabular}
\end{center}
\caption{
\label{tab:quality}
The total chi-square, $\chi^2_\text{tot}$, of the CT18A, CT18As, and CT18As\_Lat fits, respectively. The total number of data points (without including the lattice data) of each fit is 3674, and $Q_0=1.3$ GeV.
}
\end{table}


\begin{figure}[htbp]
\includegraphics[width=0.49\textwidth]{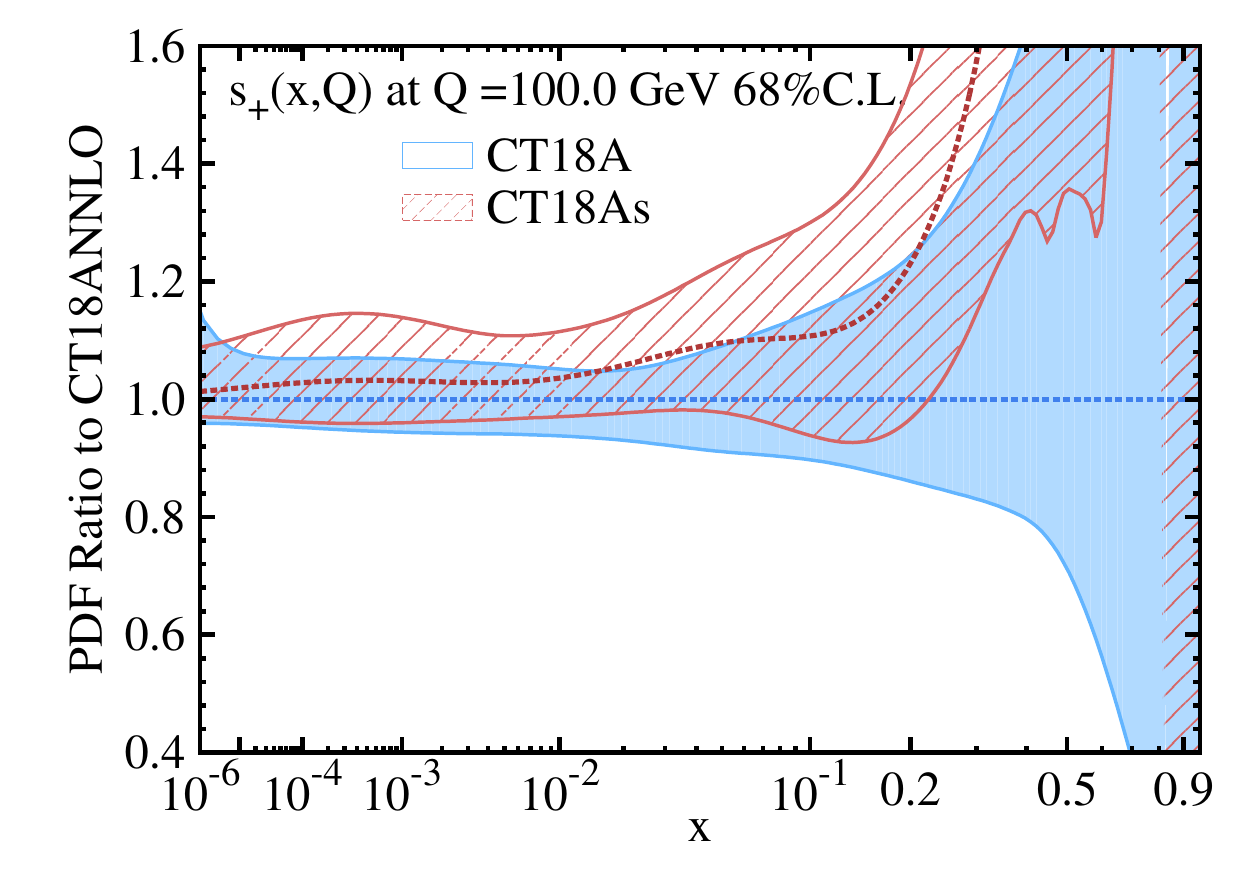}
\includegraphics[width=0.49\textwidth]{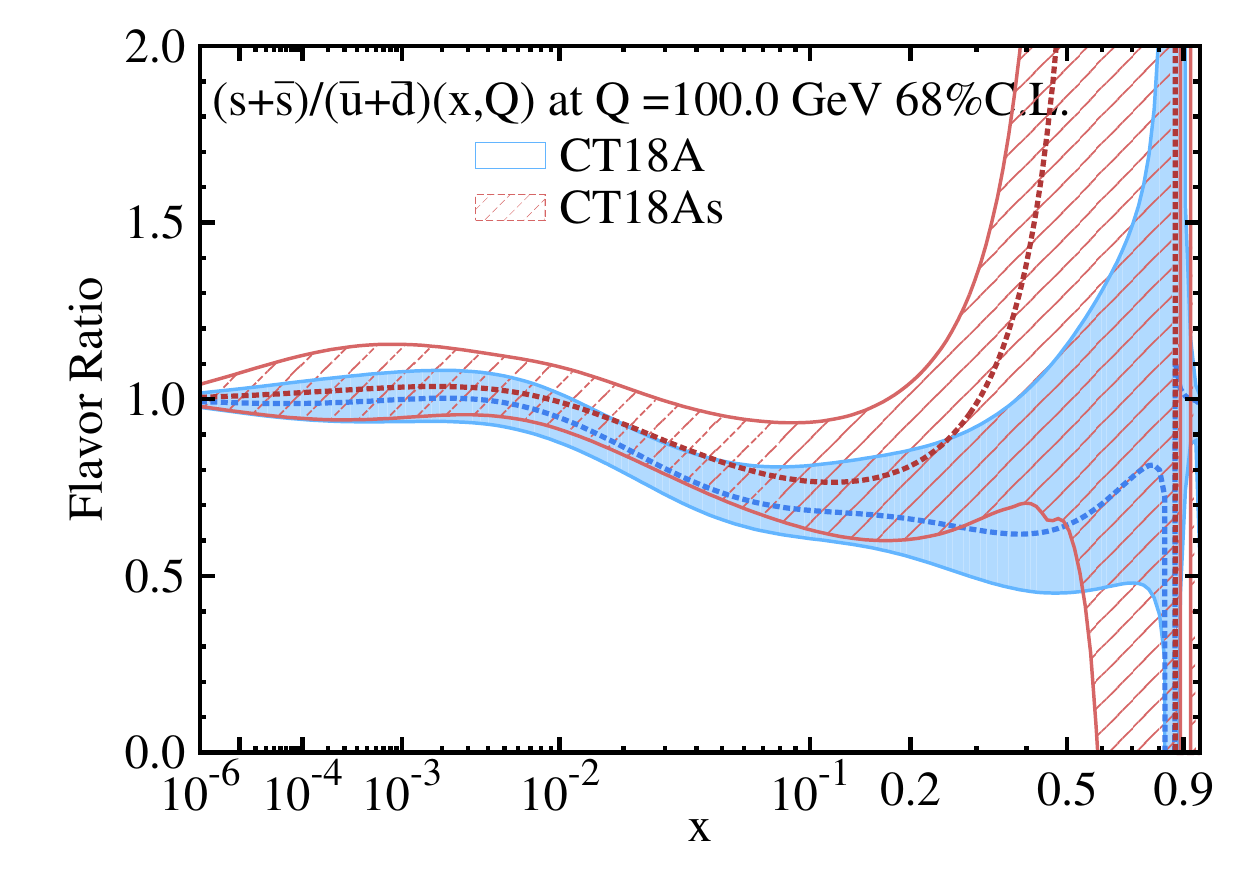}
\caption{
\label{fig:CT18Aas_1}
The comparison of $s_+ \equiv s+\bar{s}$ and $(s+\bar{s})/(\bar{u}+\bar{d})$ PDFs  for CT18A and CT18As at $Q=100$ GeV.
In CT18A, $s(x) = \bar{s}(x)$ is assumed in the parametrisation, while in CT18As the strangeness asymmetry $s_-(x) \equiv s(x) - \bar{s}(x)$ is parametrised, at the initial scale $Q_0=1.3$ GeV.}
\end{figure}

In Fig. \ref{fig:CT18Aas_1}, we compare the CT18A and CT18As for $s_+$ and $(s+\bar{s})/(\bar{u}+\bar{d})$ at $Q=100$ GeV. In CT18As, the central value of the total strangeness $s_+$ is enhanced across a wide range of $x$ from the $s_+$ distribution in CT18A. The uncertainty of $s_+$ in CT18As is also enlarged. 
The similar behaviour can also be observed in the ratio of total strangeness and light quarks $(s+\bar{s})/(\bar{u}+\bar{d})$.

\begin{figure}[htbp]
	\includegraphics[width=0.49\textwidth]{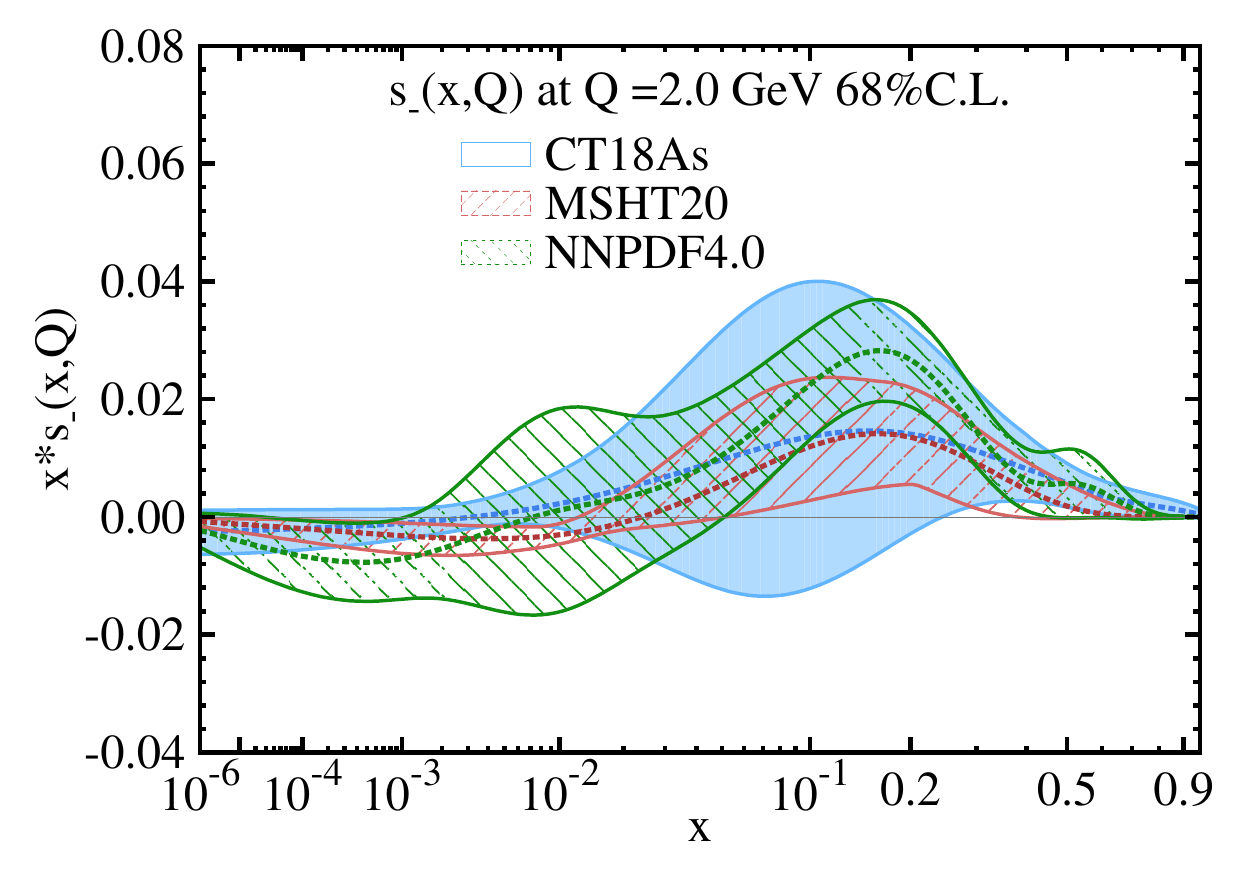}
	\includegraphics[width=0.49\textwidth]{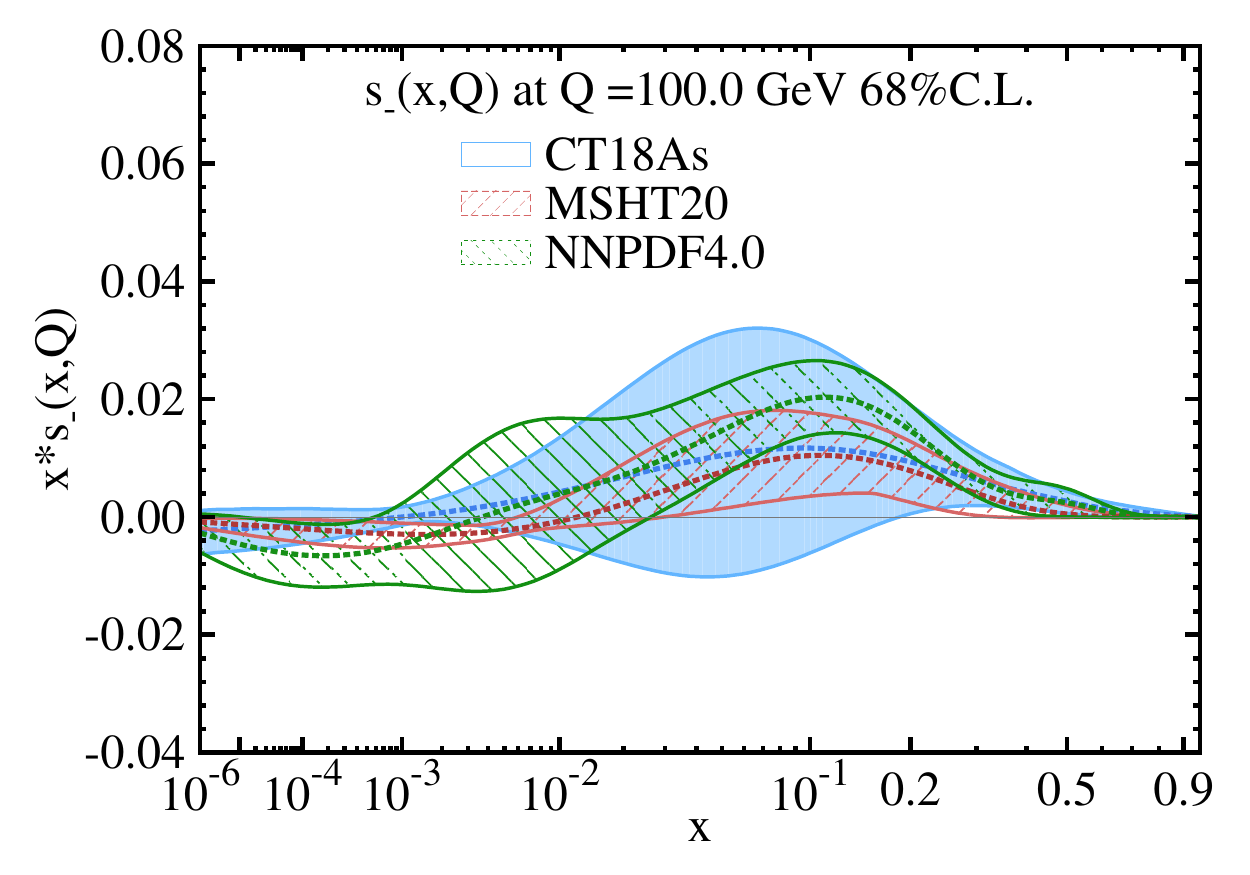}
	\caption{
	\label{fig:CT18Aas_Lat_other_PDF}
	The $s_-(x)$ distributions at 2 GeV (left) and 100 GeV (right) of CT18As are compared to those of MSHT20 \cite{Bailey:2020ooq} and NNPDF4.0 \cite{Ball:2021leu}.
}
\end{figure}

In Fig. \ref{fig:CT18Aas_Lat_other_PDF}, the $s_-(x)$ distributions at 2.0 GeV and 100 GeV of CT18As are compared to PDF fitting results by other groups. The CT18As agrees with MSHT20 \cite{Bailey:2020ooq} in terms of $s_-$ central values. For $x \sim 0.1$, NNPDF4.0 \cite{Ball:2021leu} presents the largest $s_-$ central value. In the range of $0.05 < x < 0.4$, CT18As shows a wide error band, so that CT18As is consistent with $s_-$ PDF obtained by other groups.

\begin{figure}[htbp]
\includegraphics[width=0.6\textwidth]{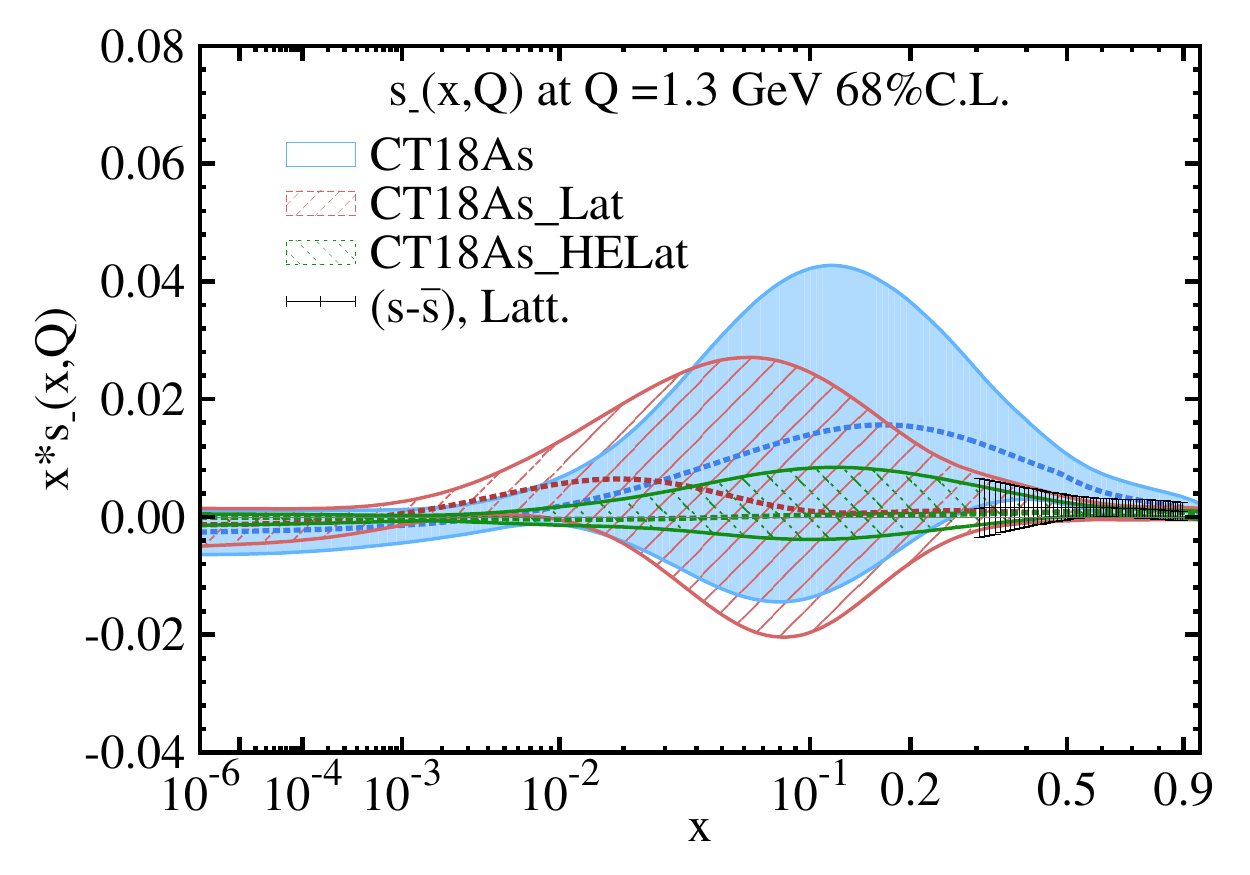}
\caption{
\label{fig:CT18Aas_Lat_Predicted} 
The result of $xs_-(x, Q=1.3 \, {\rm GeV})$ from the original CT18As fit  (blue band), with current lattice constraints (red slashed area), and expected improvement if current lattice data errors are reduced by a half (green backslashed area); the black bars are the current lattice data.  
}
\end{figure}

The impact of the lattice data on the determination of $s_-(x)$ at $Q=1.3$ GeV is shown in Fig. \ref{fig:CT18Aas_Lat_Predicted}. The lattice data points distribute in the region of $x>0.3$. Comparing to the error band of CT18As, the uncertainty in lattice data points is quite small, so that including the lattice data in the CT18As\_Lat fit greatly reduces the $s_-$-PDF error band size in the large $x$ region. The amount of reduction of the CT18As\_Lat error band into the much smaller $x$ region is likely to depend on the chosen nonperturbative parametrization form of $s_-(x)$ at $Q_0=1.3$ GeV. Hence, it is important to have more precise lattice data, extended to smaller $x$ values.

Based on the CT18As\_Lat PDF, we further investigate how much a lattice data with higher precision is able to constrain the $s_-$ distribution. We again fit the lattice data, but reduce the uncertainty of lattice data points by half, resulting another PDF labelled ``CT18As\_HELat". The half-error lattice data shows a strong power in further constraining $s_-$ by reducing the error band of $s_-$ by nearly a factor of two in the large $x$ region.

\section{Conclusion} 
\label{conclusion}

In this work, we study the impact of the lattice data on the determination of the strangeness asymmetry distribution $s_-(x) \equiv s(x) - {\bar s}(x)$ in the general CTEQ-TEA global analysis of parton distribution functions (PDFs) of the proton.
We start with the CT18A NNLO fit~\cite{Hou:2019efy}, rather than the nominal CT18 NNLO fit, since the tensions between the precision ATLAS $\sqrt{s} = 7$ TeV $W$, $Z$ data~\cite{ATLAS:2016nqi} and NuTev~\cite{Mason:2006qa} and CCFR~\cite{NuTeV:2001dfo} DIS dimuon data can be released by introducing $s(x) \neq \bar{s}(x)$, and that the mentioned ATLAS data is included in the CT18A fit and absent in the CT18 fit.
We extend the non-perturbative parametrisation in the CT18A analysis by allowing a strangeness asymmetry distribution $s_-(x) \equiv s(x) - \bar{s}(x)$ at the initial $Q_0$ scale. The resulting PDF set from the CT18A data set is labelled as CT18As, whose quality of fit is similar to the CT18A fit.
The constraint from the lattice data into the PDF global fit is added by using the Lagrange Multiplier method. We found that the resulting PDF, named as CT18As\_Lat, present a different strangeness asymmetry distribution and a smaller uncertainty band than those of CT18As. We also investigate the possible constraint of the lattice data with higher precision by performing a PDF fit with errors in the original lattice data points reduced by half. 
Our results conclude that the current lattice data is able to help constraining the strange asymmetry $s_-(x)$ in PDF global analysis. Further precision improvement in the lattice errors of this quantity can further improve the $s_-(x)$ to $x \in [10^{-2},0.6]$


\section{Acknowledgment}
This work is partially supported by the U.S. National Science Foundation
under Grant No.PHY-1653405 and PHY-2013791.  
C.-P.~Yuan is also grateful for the support from
the Wu-Ki Tung endowed chair in particle physics.


\bibliographystyle{utphys}
\bibliography{CT18As_Lat_snowmass}


\end{document}